%% file: main.tex
\documentclass[%
 reprint,
 amsmath,amssymb,
 aps,
floatfix,
]{revtex4-2}

\usepackage{graphicx}%
\usepackage{dcolumn}%
\usepackage{bm}%

\usepackage{graphicx}
\usepackage[colorlinks=true, allcolors=blue]{hyperref}
\usepackage{booktabs}
\usepackage{subcaption}
\usepackage{xcolor}

\begin{document}
\title{OTCliM: generating a near-surface climatology of \\optical turbulence strength ($C_n^2$) using gradient boosting}
\author{Maximilian Pierzyna}
\affiliation{Department of Geoscience and Remote Sensing, Delft University of Technology, Delft, The Netherlands}
\author{Sukanta Basu}
\affiliation{Atmospheric Sciences Research Center, University at Albany, Albany, USA}
\altaffiliation[Also at ]{Department of Environmental and Sustainable Engineering, University at Albany, Albany, USA}
\author{Rudolf Saathof}
\affiliation{Faculty of Aerospace Engineering, Delft University of Technology, Delft, The Netherlands}

\date{\today}

\begin{abstract}
  This study introduces OTCliM (Optical Turbulence Climatology using Machine learning), a novel approach for deriving comprehensive climatologies of atmospheric optical turbulence strength ($C_n^2$) using gradient boosting machines. 
  OTCliM addresses the challenge of efficiently obtaining reliable site-specific $C_n^2$ climatologies near the surface, crucial for ground-based astronomy and free-space optical communication.
  Using gradient boosting machines and global reanalysis data, OTCliM extrapolates one year of measured $C_n^2$ into a multi-year time series.
  We assess OTCliM's performance using $C_n^2$ data from 17 diverse stations in New York State, evaluating temporal extrapolation capabilities and geographical generalization.
  Our results demonstrate accurate predictions of four held-out years of $C_n^2$ across various sites, including complex urban environments, outperforming traditional analytical models. 
  Non-urban models also show good geographical generalization compared to urban models, which capture non-general site-specific dependencies. 
  A feature importance analysis confirms the physical consistency of the trained models.
  It also indicates the potential to uncover new insights into the physical processes governing $C_n^2$ from data.
  OTCliM's ability to derive reliable $C_n^2$ climatologies from just one year of observations can potentially reduce resources required for future site surveys or enable studies for additional sites with the same resources.
\end{abstract}

\maketitle

\section{Introduction}
Atmospheric optical turbulence is highly relevant for optical ground-based astronomy and future free-space optical communication (FSOC).
Both applications suffer from light getting distorted when propagating through the turbulent atmosphere.
In astronomy, turbulent fluctuations of the atmospheric refractive index, known as optical turbulence (OT), cause blurry images and limit the detection of small objects \citep{hardy1998}.
FSOC links, which use optical beams to transmit data instead of traditional radio waves, experience reduced data rates or even link interruptions due to OT \citep{kaushal2017,jahid2022}.
Therefore, the optical turbulence strength ($C_n^2$, where index $n$ denotes the refractive index) must be carefully considered in the design and operation of such optical systems.
This requires robust statistical evaluation of $C_n^2$ over time for the operational sites of interest, known as a $C_n^2$ climatology.
Such a climatology, derived from long-term $C_n^2$ data, portrays trends, seasonal variations, or potential anomalies in OT strength.

Obtaining $C_n^2$ climatologies is challenging because turbulence strongly depends on the local topography and varying meteorological conditions.
That is why long-term on-site surveys measuring local OT conditions are crucial.
For instance, before a new optical observatory is built, OT strength is typically measured for multiple years at a few carefully selected locations to identify the best one (see, e.g., \citet{hill2006,schock2009}). 
When envisioning a global FSOC network, site surveys have so far been focused on cloud cover (see, e.g., \citet{fuchs2015,poulenard2015,pham2023}), but similar site surveys targeting OT are expected to become highly relevant in the future.

However, conducting long-term site surveys at various locations of interest is time-consuming and resource-intensive.
While $C_n^2$ can be obtained by post-processing numerical weather model output (e.g., \citet{masciadri1999}), running such models for multiple years is computationally expensive, and the accuracy of the resulting $C_n^2$ is very sensitive to the model configuration \citep{pierzyna2023b} and the selected $C_n^2$ parameterization \citep{pierzyna2024}.
We address these issues by proposing a novel machine learning-based (ML) approach called OTCliM (Optical Turbulence Climatology using Machine learning).
OTCliM aims to extrapolate just one year of measured near-surface $C_n^2$ into a multi-year time series, enabling the generation of a comprehensive site-specific $C_n^2$ climatology with less data than a conventional site survey.
Our approach does not yet provide full vertical $C_n^2$ profiles as often sought in astronomy or FSOC.
But for near-surface conditions, which are highly relevant for, e.g., horizontal near-surface FSOC links,
site survey costs can potentially be reduced, climatologies can be obtained faster, and more sites can be surveyed within a given time frame.
By leveraging ML, OTCliM can also model complex input-output relations, offering an advantage over traditional empirical $C_n^2$ models, which are discussed in the following section.

This study assesses OTCliM's performance using an extensive $C_n^2$ dataset containing measurements from 17 diverse stations. 
We evaluate the temporal extrapolation capabilities of the trained models and analyze their potential for geographical generalization to other sites.
Additionally, we examine the importance of each input variable for predicting $C_n^2$ to probe if the ML models have learned physically plausible relations.
That analysis increases confidence in the ML models and could uncover new dependencies of $C_n^2$ from the data.

\section{Relevant $C_n^2$ regression studies}
The experimental determination of $C_n^2$ is challenging because it requires expensive instruments, such as scintillometers \citep{beyrich2021} or meticulous post-processing of high-frequency temperature measurements \citep{beason2024}.
Consequently, there has been a sustained effort over several decades to develop parameterizations of $C_n^2$ using more readily available meteorological variables.
Conventional approaches to $C_n^2$ parameterization typically involve the application of physics-based analytical equations in conjunction with empirically derived coefficients or regression functions.
A popular example is the formulation proposed by \citet{wyngaard1971}, based on Monin-Obukhov similarity theory (MOST) \citep{monin1954}.
That parameterization -- called W71 in the following -- relates sensible heatflux $\overline{w'\theta'}$ and friction velocity $u_*$ to the strength of temperature fluctuations ($C_T^2$) at height $z$ as
\begin{equation}
    C_T^2=\left(-\overline{w'\theta'}/u_*\right)^2 z^{-2/3} g(\zeta).
    \label{eq:w71}
\end{equation}
Sensible heat flux $\overline{w'\theta'}$ and friction velocity $u_* = {\left(\overline{u'w'}^2 + \overline{v'w'}^2\right)}^{1/4}$ describe the vertical (i.e., surface-normal) turbulent transport of heat ($\theta'$) and momentum ($u'$, $v'$) due to fluctuations of the vertical wind component ($w'$) \citep{stull1988}.
As these parameters capture the two effects modulating turbulence -- buoyancy (sensible heat flux) and wind shear (friction velocity) --, they are commonly used in turbulence parameterizations.
Since refractive index fluctuations ($C_n^2$) are driven by density fluctuations due to temperature and moisture fluctuations, $C_n^2$ is linked to $C_T^2$ as \citep{wesely1976,moene2003}
\begin{equation}
  C_n^2 = {\left(A\, \frac{P}{\overline{T}^2}\right)}^2 {\left(1 + \frac{0.03}{\beta}\right)}^2\, C_T^2
  \label{eq:gladstone}
\end{equation}
with $A \approx 7.9\times10^{-5}\, \text{K}\, \text{hPa}^{-1}$ for optical wavelengths \citep{andreas1988},
station pressure $P$ in hPa, and mean air temperature $\overline{T}$ in K.
The contribution of moisture fluctuations to $C_n^2$ is captured by the Bowen ratio $\beta$, the ratio of the sensible heat flux and the latent heat flux due to evaporation.
If $C_n^2$ is estimated above the boundary layer in the free atmosphere, a modified formulation of Eq.~\ref{eq:gladstone} can be used \citep{cherubini2013}.

Function $g(\zeta)$ in Eq.~\ref{eq:w71} is a similarity function, empirically determined from observations,
\begin{equation}
    \begin{split}
        g(\zeta) &= 4.9 {(1 - 6.1 \zeta)}^{-2/3}, \zeta < 0 \\
        g(\zeta) &= 4.9 (1 + 2.2 \zeta^{2/3}), \zeta \geq 0
    \end{split}
    \label{eq:w71_sim_func}
\end{equation}
with stability parameter $\zeta = z/L$ where $L$ is the Obukhov length $L = - u_*^3\, \overline{T} / (\kappa\, g\, \overline{w'\theta'})$ and $z$ the height above ground.
In $L$, 
$g=9.81$\,{m\,s$^{-2}$} is the gravitational acceleration of Earth and
$\kappa=0.4$ is the Von Kármán constant.
Various alternative regression-based similarity functions $g(\zeta)$ were proposed in literature aimed at improving the accuracy and applicability of $C_n^2$ estimates across diverse meteorological and topographic conditions (see, e.g., \citet{savage2009} for an extensive review).

In recent years, different machine learning (ML) techniques have been utilized to derive $C_n^2$ parameterizations.
Similar to W71, ML-based approaches aim to obtain regression models that estimate $C_n^2$ from routine meteorological variables (temperature, pressure, wind speed), gradients (potential temperature gradient or wind shear), or heat and radiation fluxes. 
However, instead of deriving physics-based functional expressions (cf.\ Eqns.~\ref{eq:w71}~and~\ref{eq:gladstone}), which are fitted to observations (cf.\ Eq.~\ref{eq:w71_sim_func}), ML directly models the relation between observed $C_n^2$ and the input variables.
The power of such ML regression models is that they can model complex multi-variate relations, yielding potentially better $C_n^2$ estimates compared to traditional approaches.

\begin{figure*}[t]
  \centering
  \includegraphics[width=\textwidth]{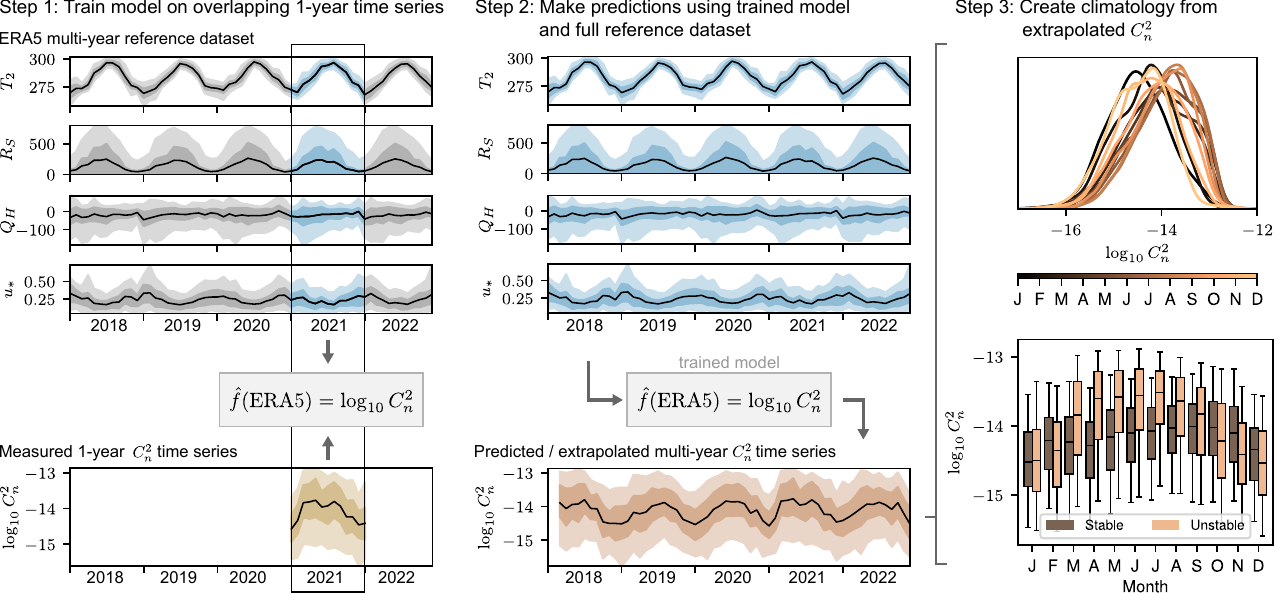}
  \caption{Proposed OTCliM approach to extrapolate a measured 1-year time series of optical turbulence strength (golden yellow) to multiple years (orange) based on ERA5 reference data (blue). Robust yearly $C_n^2$ statistics can be obtained from the extrapolated data.}%
  \label{fig:mcp}
\end{figure*}

The ML-based $C_n^2$ parameterizations in literature differ in their input variables, 
the ML regression technique employed, and 
if $C_n^2$ is measured at a single or multiple vertical levels.
Much work has been devoted to single-level near-surface $C_n^2$ obtained as point measurements (e.g., from sonic anemometer) or via path-averaging (e.g., from scintillometer).
\Citet{wang2016}, for example, used fully connected neural networks to estimate single-level $C_n^2$ from routine meteorological data and gradients.
\Citet{vorontsov2020} also utilized neural networks but aimed at estimating path-averaged $C_n^2$ from images of a received laser beam that is broken up into speckles due to turbulence. 
\Citet{jellen2020,jellen2021} employed random forest and gradient boosting machines (GBM) to estimate path-averaged $C_n^2$ in a maritime environment from routine parameters and radiation fluxes.
Most ML-based models do not yield analytical equations of the fitted model, 
unless they explicitly target it as shown by \citet{bolbasova2021}.
The authors obtained a complex site-specific equation for single-level $C_n^2$, similar to non-ML empirical models proposed by, e.g., \citet{sadot1992}, \citet{wang2015}, or \citet{arockiabazilraj2015}.
The difference between these empirical models and MOST-based models is that they are not derived physically but aim to capture the local optical turbulence strength well.
Going beyond single-level $C_n^2$ estimates, \citet{pierzyna2023a} proposed a physics-inspired framework to derive a data-driven similarity theory of optical turbulence that combines $C_n^2$ observations from multiple levels within the surface layer into a single non-dimensional model.
In contrast to the near-surface studies, \citet{su2021} proposed a method to estimate a vertical $C_n^2$ profile by modeling OT strength as a function of radiosonde measurements.
\Citet{cherubini2021} presented an ML framework to estimate seeing, which is the vertical integral of the $C_n^2$ profile, close to the surface and in the free atmosphere, but does not explicitly resolve vertical levels.
\Citet{milli2020} tackled yet another estimation angle by using ML to forecast future OT conditions for the next two hours based on past observations and auxiliary features.
However, all these approaches require in-situ data, rendering them unsuitable for temporal extrapolation as targeted in the present study.
Additionally, in-situ-based models need sensors deployed at the site of interest for $C_n^2$ estimation, which hinders their applicability to sites without instrumentation.
To alleviate this issue, we proposed OTCliM, which parameterizes $C_n^2$ based on reanalysis data, which are available globally and for multiple decades into the past.

\section{Methodology}\label{sec:meth}

Our proposed OTCliM\footnote{The Python code for training is available on GitHub: \url{https://github.com/mpierzyna/otclim}} approach aims to extrapolate one year of observed $C_n^2$ to multiple years to obtain a robust statistical description of OT strength at a particular site.
An overview of OTCliM is given in Fig.~\ref{fig:mcp}, which we base on the measure-correlate-predict (MCP) framework popular in wind energy \citep{carta2013,kartal2023}.
In the first step, $C_n^2$ is measured for one year (golden yellow) and then correlated to a reference dataset.
For OTCliM, we utilize variables from the ERA5 reanalysis \citep{era5} extracted at the location of the $C_n^2$ measurements as reference data and gradient boosting machines (GBM) to regress the ERA5 time series to $C_n^2$ where they overlap (blue).
In the second step, the trained model is utilized to extrapolate $C_n^2$ based on ERA5 to multiple years (prediction step in MCP), which can then be used to obtain site-specific seasonal statistics as presented in step 3.

We utilize GBMs for the regression step (cf.~\S\ref{sec:meth_gb}) because they are known for their high performance in non-linear tabular regression tasks.
For comparison, we also train GBM models that use in-situ observations instead of ERA5 as input data, which serve as performance baselines for OTCliM.
Similarly, the traditional W71 model is evaluated with ERA5 inputs to provide a second baseline (cf.~\S\ref{sec:meth_baseline}).
To ensure our GBM models capture physically reasonable dependencies, we quantify the feature importance assigned to each ERA5 input variable (cf.~\S\ref{sec:meth_shap}). 
Finally, the performance of the trained models is assessed in terms of their temporal extrapolation and geographical generalization capabilities (cf.~\S\ref{sec:meth_eval}).

\subsection{Gradient boosting regression}\label{sec:meth_gb}
Following the MCP framework presented in Fig.~\ref{fig:mcp}, we aim to train one ML model for one site using training data from one year.
The corresponding training data are the input data $\boldsymbol{X}_{s,t}\in\mathbb{R}^{n\times p}$ 
and the target vector $\boldsymbol{y}_{s,t} \in \mathbb{R}^n$,
where $s \in \mathcal{S}$ is one out of multiple sites $\mathcal{S}$ 
and $t \in \mathcal{T}$ is one out multiple training years $\mathcal{T}$ available at that site.
The ML task is to learn a regression function $\hat{f}_{s,t}$ that approximates a sample $y_i \in \boldsymbol{y}_{s,t}$ from the target vector based on a sample $\boldsymbol{x}_i \in \boldsymbol{X}_{s,t}$ of the input data:
$\hat{f}_{s,t}(\boldsymbol{x}_i) = \hat{y}_i \approx y_i$.
$\boldsymbol{X}_{s,t}$ and $\boldsymbol{y}_{s,t}$ contain a matching number of $n$ samples, i.e., the samples of the one-year time series, 
where $\boldsymbol{X}_{s,t}$ is composed of $p$ features, i.e., concurrent time series of different meteorological variables, 
and $\boldsymbol{y}_{s,t}$ contains the scaled $\log_{10}C_n^2$.

The OT strength $C_n^2$ is $\log_{10}$-transformed because it varies over multiple orders of magnitude throughout the day, which is challenging to capture in non-log space.
Also, the range of yearly $C_n^2$ variation varies between sites, so $\log_{10}C_n^2$ is scaled before training to make the performance scores and feature importance values comparable between sites. 
More details about the $C_n^2$ pre-processing are presented in \S\ref{sec:data_nysm}.

The GBM regression models $\hat{f}_{s,t}$ are trained using the AutoML library FLAML \citep{wang2021}.
FLAML is a time-constrained hyperparameter optimizer that aims to find the optimal GBM model configuration within a specified time budget. 
FLAML optimizes not only the hyperparameters of a single algorithm but also explores switching between algorithms, here, the two popular algorithms XGBoost \citep{chen2016} and LightGBM \citep{ke2017}.
The internal loss functions of XGBoost and LightGBM are also treated as hyperparameters to be optimized by FLAML with $L_1$-norm, $L_2$-norm, and Huber norm as options.
The hyperparameter optimization uses 5-fold cross-validation based on the training data and aims to minimize the root-mean-squared error between the predicted and hold-out $\boldsymbol{y}$.
Each model is trained for 45\,min on 8 CPU cores of an Intel Xeon 2648R, resulting in 6 core hours per model.

\subsection{Baseline models}\label{sec:meth_baseline}
We employ two baseline models to put the performance of OTCliM into perspective.
The first model is the traditional empirical W71 model as given by Eqns.~\ref{eq:w71} and \ref{eq:w71_sim_func}.
Since the similarity function $g(\zeta)$ does not contain site-specific model coefficients, the accuracy of the W71 $C_n^2$ estimates is expected to be lower than the site-specific OTCliM models.
Secondly, we utilize GBM models trained with in-situ observations as input instead of reanalysis.
Reanalysis datasets are coarse-resolution model outputs (spatial resolution typically $\mathcal{O}(10\,\text{km})$), which are expected to miss some local and complex processes and patterns modulating turbulence.
Such processes would be contained in the in-situ data, so the performance of in-situ GBM models is expected to be higher than that of ERA5-based OTCliM models.
These baselines allow us to disentangle the influence of the modeling approach (traditional vs.\ GBM) from that of the input data (ERA5 vs.\ in-situ).
Appendix~\ref{app:baseline} gives more technical details about both baseline models.

\subsection{Physical consistency checking using feature importance}\label{sec:meth_shap}
To assess if the trained models captured physically consistent relationships and to potentially discover currently unknown dependencies between features and the regression target, we quantify the importance of each feature for the prediction using SHapely Additive exPlanation (SHAP) values \citep{lundberg2017}.
SHAP values explain how a trained model $\hat{f}_{s,t}$ arrives at its predictions $\hat{\boldsymbol{y}}$ on a per-sample basis.
For each sample $\boldsymbol{x}_i$, the prediction $\hat{y}_i$ is explained as
\begin{equation}
  \hat{f}_{s,t}(\boldsymbol{x}_i) = \hat{y}_i = \mathbb{E}\left[ \hat{\boldsymbol{y}}_i \right] + \sum_{j=1}^p \phi_{j,i}
  \label{eq:shap_values}
\end{equation}
where $\phi_{j,i}$ are the SHAP values of the $i$-th sample and $\mathbb{E}\left[ \hat{\boldsymbol{y}}_i \right]$ is the expected value of all predictions.
In plain words, the $j$-th SHAP value $\phi_{j,i}$ describes the contribution of the $j$-th feature to the prediction assuming a local linear model.
A large step $\phi_{j,i}$ compared to the other SHAP values indicates that the $j$-th feature contributes strongly to that prediction and must be important. 
Consequently, a global view of the importance of the $j$-th feature can be obtained by averaging the $j$-th SHAP values for all $n$ samples of the dataset \citep{molnar2022} such that
\begin{equation}
  I_j = \frac{1}{n}\sum_{i=1}^{n}\lvert\phi_{j,i} \rvert.
  \label{eq:shap_fi}
\end{equation}

Note that SHAP values explain a feature's contribution to a model's prediction $\hat{y}_i$, not to the true value $y_i$.
Accurate feature importance (FI) estimates, thus, require well-performing models $\hat{f}{s,t}$ where $\hat{y}_i$ closely approximates $y_i$ for training and testing data.
As predictions typically align more closely with true values in training data than testing data, we compute SHAP values using the former. 
This approach allows us to examine the model's internal workings without the influence of the generalization error. 
Moreover, it aligns with practical scenarios where all available data is used for training the final model, meaning that the FI analysis must be performed using the training dataset \citep{molnar2022}.

\subsection{Performance evaluation}\label{sec:meth_eval}
The performance of the trained OTCliM models is quantified through the Pearson correlation coefficient $r$ and the root-mean-squared error (RMSE) $\epsilon$:
\begin{equation}
  r=\frac{\sum_{i=1}^n(y_i - \overline{y})(\hat{y}_i - \overline{\hat{y}})}{n\, \sigma_y\, \sigma_{\hat{y}}},
  \label{eq:met_r}
\end{equation}
\begin{equation}
  \epsilon = \sqrt{\frac{1}{n} \sum_{i=1}^n {\left(y_i - \hat{y}_i\right)}^2}.
\end{equation}
In Eq.~\ref{eq:met_r}, the overbar denotes the mean and $\sigma$ the standard deviation of true ($\boldsymbol{y}$) and estimated ($\hat{\boldsymbol{y}}$) scaled $C_n^2$, respectively.
Two validation strategies are employed to assess the temporal and geographical extrapolation capabilities of the OTCliM models.
Note that only the testing data used to compute the scores differ between the evaluation strategies -- the training data and the trained model remain the same.

\paragraph{Temporal extrapolation}
For the MCP application -- the temporal extrapolation --, the model $\hat{f}_{s,t}$ trained on data from year $t \in \mathcal{T}$ needs to accurately predict OT strength at the same site for the held-out testing years $\mathcal{T}\setminus t$: $\hat{f}_{s,t}(\boldsymbol{x}_i) = \hat{y}_i$ with $\boldsymbol{x}_i \in \boldsymbol{X}_{s,\mathcal{T}\setminus t}$ and $\hat{y} \in \boldsymbol{\hat{y}}_{s,\mathcal{T} \setminus t}$.
The scores are computed between the true held-out data $\boldsymbol{y}_{s,\mathcal{T} \setminus t}$ and the predictions $\boldsymbol{\hat{y}}_{s,\mathcal{T} \setminus t}$ and are denoted $\square_{s,t}$, where $\square$ is a placeholder for the performance metrics $r$ and $\epsilon$.
Training happens in a round-robin fashion, so one model is trained per training year and evaluated for the hold-out years.
This process is repeated until all years are used for training once.
The average performance of these models for a specific station $s$, called the MCP score, is obtained by averaging the scores across all training years: $\overline{\square}_s = {\langle\square_{s,t}\rangle}_{t\in\mathcal{T}}$.

\paragraph{Cross-site evaluation}
The second evaluation strategy probes the geographical generalization capability of a model trained on station $s$ when applied to another station $\tilde{s} \in \mathcal{S} \setminus s$.
The trained model is tasked to predict OT strength at $\tilde{s}$ given the full multi-year input data $\boldsymbol{X}_{\tilde{s},\mathcal{T}}$ at $\tilde{s}$ 
from which the cross-site scores using $r$ and RMSE can be obtained: $\square_{(s,t) \rightarrow (\tilde{s},\mathcal{T})}$. 
For readability, we average the cross-site scores achieved by the individual models per training site $s$: $\square_{s \rightarrow \tilde{s}} = {\langle\square_{(s,t)\rightarrow(\tilde{s},\mathcal{T})}\rangle}_{t \in \mathcal{T}}$.

\section{Dataset}\label{sec:data}
To evaluate OTCliM's performance, we train GBM models for 17 diverse locations across New York State (cf.~\S\ref{sec:data_nysm}).
Five years of $C_n^2$ are available at each site for which we select collocated and concurrent ERA5 data (cf.~\S\ref{sec:data_era5}).

\subsection{New York State Mesonet (NYSM)}\label{sec:data_nysm}
\begin{table}[t]
  \centering
  \caption{Flux stations of the New York State Mesonet used to benchmark the OTCliM approach. The three urban stations are marked with (*).}%
  \label{tab:nysm_names}
  \resizebox{\columnwidth}{!}{\input{nysm_names.tex}}
\end{table}

The New York State Mesonet (NYSM) comprises 127 standard weather stations as of 2024 spread across New York State, United States of America, and has been fully operational since 2018 \citep{brotzge2020}.
These weather stations measure routine meteorological parameters such as 2\,m-temperature, 10\,m-wind speed and direction, surface pressure, and several other variables. 
The sampling rate of these measurements is on the order of seconds, and final values are reported as 10\,min aggregates (mean and variance).
However, measuring $C_n^2$ requires higher-frequency observations to resolve the inertial range of turbulence.
Such high-frequency measurements are available at a subset of 17 stations -- the flux stations -- additionally equipped with Campbell Scientific CSAT3A sonic anemometers mounted at 9\,m height.
The NYSM sonic anemometers measure the three wind components and the sonic temperature at $f_s=10$\,Hz \citep{brotzge2020}, which is high enough to obtain $C_n^2$.

We utilize the data from these NYSM flux stations to obtain training targets for the OTCliM models.
The flux stations are placed in diverse topographical and climatological environments as listed in Tab.~\ref{tab:nysm_names}.
The stations BKLN, QUEE, and STAT are located on rooftops in urban environments where measurements are strongly influenced by their immediate surroundings.
Neighboring buildings can, for example, cast shadows or cause wakes, which influence radiation, wind, and, therefore, also the local turbulence \citep{wmo8iii}.
Since we expect these urban stations to behave differently than the rural stations, they are marked with (*) in the table. 
For each of the 17 stations, five years of measurements are available.
Following the notation introduced previously, the set of flux stations is denoted $\mathcal{S}$, and $\mathcal{T}$ represents the set of five training years 2018 -- 2022 available at each site.
The corresponding target vector $\boldsymbol{y}_{s,t}$ is obtained by estimating $C_n^2$ from sonic anemometer measurements, applying quality assurance and control (QA/QC) steps, and scaling the OT data to make them comparable between different sites.
These stages are described in detail below.

\subsubsection{Structure function approach}
As noted in the context of Eq.~\ref{eq:gladstone}, $C_n^2$ quantifies the strength of refractive index fluctuations due to density fluctuations caused by turbulent temperature and moisture fluctuations.
The temperature $T_s$ measured by sonic anemometers also contains a humidity contribution as $T_s = T(1+0.51q)$ with specific humidity $q$ \citep{kaimal1991}.
We assume that using the simplified version
\begin{equation}
  C_n^2 \approx {\left(A\, P/\overline{T}^2\right)}^2 C_{T_s}^2
\end{equation}
of Eq.~\ref{eq:gladstone} implicitly accounts for moisture if the strength of these sonic temperature fluctuations, $C_{T_s}^2$, is used.
That is because $C_{T_s}^2$ can be decomposed as $C_{T_s}^2 \approx C_T^2 + 1.02 \overline{T} C_{Tq} + 0.26 \overline{T}^2 C_q^2$ with
$C_T^2$ and $C_q^2$ representing the strength of pure temperature and moisture fluctuations and $C_{Tq}$ the cross-term between the two.
For readability, we will refer to $C_{T_s}^2$ simply as $C_T^2$ but would like to reiterate that a moisture contribution is included implicitly.

In the inertial range, $C_T^2$ is the coefficient of the second-order structure function of the (sonic) temperature defined as
\begin{equation}
  S_{T}^2(\Delta x) = \langle{\left(T(x) - T(x+\Delta x)\right)}^2\rangle \overset{\substack{\hidewidth \Delta x\, \in\, \text{inertial range}\hidewidth\\\downarrow}}{=} C_{T}^2 \, {\Delta x}^{2/3}.
  \label{eq:sf}
\end{equation}
The structure function is computed from 5\,min non-overlapping windows of the sonic temperature signal, and
$C_{T}^2$ is obtained by fitting a $2/3$ slope to the inertial range of $S_{T}^2$ in log-log coordinates. 

\subsubsection{Quality assurance and control (QA/QC)}\label{sec:data_nysm_qaqc}
Since our GBM models are purely data-driven, the quality of the fitted model strongly depends on the quality of the training data, so we apply several QA/QC steps to obtain a clean dataset.
First, $C_T^2$ \emph{has} to be determined from the slope of the inertial range, i.e., slopes of $2/3$.
Since the inertial range is not always resolved or well captured in every 5\,min window, only $C_T^2$ values with well-fitted slopes are kept ($2/3\pm5\%$ slope, $R^2>0.95$).
A similar procedure was followed by \citet{he2016a} for the analysis of simulated data.
Second, $S_T^2$ in Eq.~\ref{eq:sf} is defined spatially as a function of separation distance $\Delta x$, but the sonic anemometer measures $T$ temporally.
Taylor's frozen turbulence assumption is commonly made to convert the temporal sonic signal into a spatial form using the horizontal mean wind speed $\overline{M}$ and $\Delta x = \overline{M}/f_s$.
However, Taylor's assumption might break down in low wind conditions, so all $C_T^2$ values for $\overline{M}<1$\,m\,s$^{-1}$ are discarded.
Finally, precipitation strongly affects sonic temperature measurements \citep{zhang2016}, so only $C_T^2$ values for dry conditions can be used.
This filter is also justified from an operational perspective because optical telescopes are typically not operated in the rain, and FSOC links are strongly attenuated by snow \citep{itur1814}.

Being an observed dataset, the sonic temperature signal contains gaps due to, e.g., power or communication outages or instrument malfunctioning \citep{nysmesonet2023}, leading to $\sim16$\% of missing values on average. 
That leaves $\sim84$\% of the 5-min non-overlapping windows gap-free and suitable for the computation of $C_T^2$.
The three QA/QC steps further reduce the average number of valid samples, leaving 
$\sim66\%$ after inertial range fitting, 
$\sim55\%$ after ensuring the applicability of the Taylor assumption, and, 
$\sim53\%$ after discarding precipitation events.
Although only half of the whole dataset is considered valid,
it still contains ca.~fifty thousand samples per site and training year spread throughout diverse meteorological conditions.
This claim is supported in Appendix~\ref{app:qaqc_stab}, where we present that the QA/QC procedure generally maintains the ratio between stable and unstable atmospheric conditions before and after QA/QC.
Thus, the final dataset is considered adequate for training the OTCliM models.

\subsubsection{Distribution scaling}\label{sec:distr_scaling}
The quality-controlled $C_T^2$ time series is converted to $C_n^2$ using the Gladstone equation (Eq.~\ref{eq:gladstone}) and then scaled.
The scaling is needed because the yearly distribution of OT strength differs between sites.
Some locations show stronger OT (shifted distribution) or a larger range of $C_n^2$ values (scaled distribution) than other sites.
To make the performance scores and SHAP FI values comparable between sites, we scale the site-specific $\log_{10} C_n^2$ before training based on the 50\% log-range such that 
\begin{equation}
  \left[\log_{10}C_n^2(t)\right]_{\text{scaled}} = \frac{
    \log_{10}C_n^2(t) - \left[\log_{10}C_n^2\right]_{p25}
  }{
    \left[\log_{10}C_n^2\right]_{p75} - \left[\log_{10}C_n^2\right]_{p25}
  }.
  \label{eq:distr_scaling}
\end{equation}
Here, the subscripts p25 and p75 indicate the 25th and 75th percentiles of the original distribution used for scaling.
More details are given in Appendix~\ref{app:distr_scaling}.

\subsection{ERA5 reanalysis data}\label{sec:data_era5}
\begin{figure}[t]
  \centering
  \begin{subfigure}[t]{0.48\textwidth}
    \caption{Land sea mask}
    \includegraphics[width=\textwidth]{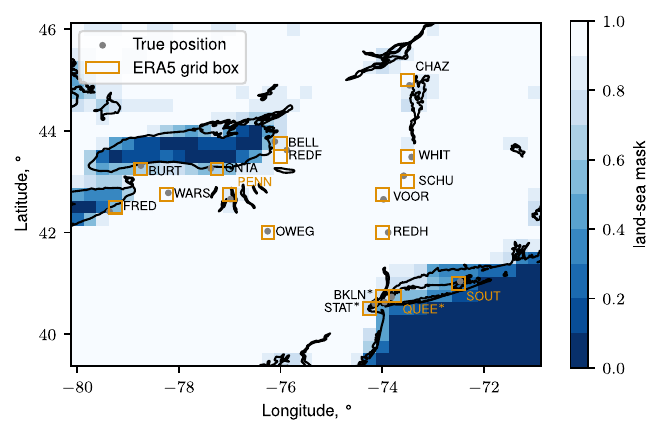}
  \end{subfigure}
  \hfill
  \begin{subfigure}[t]{0.48\textwidth}
    \caption{Geopotential height}
    \includegraphics[width=\textwidth]{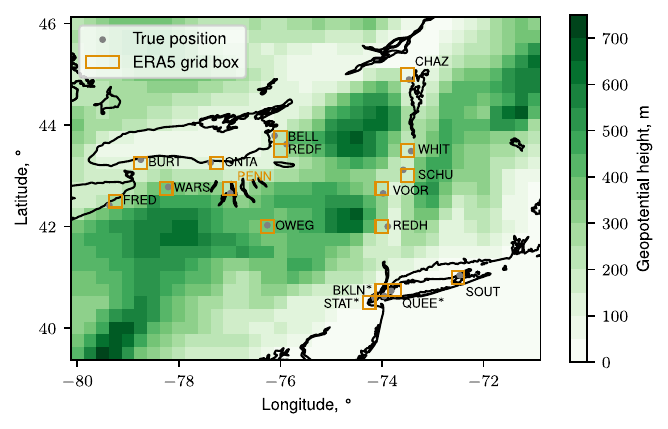}
  \end{subfigure}
  \caption{ERA5 representation of NYSM domain with true locations of NYSM flux stations (grey) and corresponding ERA5 grid box (orange) containing the stations. Urban sites are marked with (*).}%
  \label{fig:era5_domain}
\end{figure}

The ERA5 reanalysis \citep{era5} serves as the input dataset for OTCliM from which $C_n^2$ shall be estimated.
ERA5 is available globally from 1950 until the present and, thus, includes our NYSM training locations and times.
We extract ERA5 input data for all NYSM stations and training years at the ERA5 grid points closest to the respective stations.
ERA5 grid points represent 1/4-degree grid boxes (ca.\ 30\,km by 30\,km), so by selecting the closest point, we obtain data representative for the grid box containing the station.
Figure~\ref{fig:era5_domain} depicts the location of the flux stations (grey markers) and their collocated grid boxes (orange square) and also gives an impression of how the coarse ERA5 represents (a) land, sea, and (b) terrain.
The temporal resolution/sampling rate also differs between ERA5 (1\,h) and the observed $C_n^2$ dataset (5\,min). 
ERA5 data are instantaneous snapshots available at each full hour, so we match each ERA5 sample with the average of the six 5-min $C_n^2$ values $\pm 15$\,min around the hour.
This 30\,min average centered around the hour smoothes the $C_n^2$ signal to counteract potential temporal misalignment between observations and reanalysis.

\begin{table*}[t]
  \caption{ERA5 variables and variables derived thereof serve as features for the OTCliM approach. Derived features do not have ERA5 variable names and are marked with (-). }%
  \label{tab:ra_variables}
  \resizebox{\textwidth}{!}{\input{era5_vars.tex}}
\end{table*}

We aim to incorporate all variables linked to the two processes driving atmospheric turbulence: wind shear and buoyancy.
Commonly used variables include sensible heat flux and friction velocity, as utilized in W71. 
However, an advantage of using ML over deriving analytical equations from theory is the ability to include variables that may indirectly influence turbulence. 
For example, the ERA5 gravity wave dissipation rate (GWD) could be significant in complex mountainous regions where orographic gravity wave drag is known to modulate momentum fluxes \citep{lilly1972,palmer1986}. 
If there is a relationship between GWD and $C_n^2$, ML models will likely identify and utilize it, potentially revealing new dependencies.

A complete list of the ERA5 variables selected as features is presented in Tab.~\ref{tab:ra_variables}. 
Many listed features are (partially) redundant and/or (partially) correlated.
Since it is initially unknown which features are most suitable for estimating near-surface $C_n^2$, we aim for a broad feature space.
During training, the GBM algorithm will identify and base its predictions on the most important features.
Through the post-training feature importance analysis, we can identify which features the trained model deems relevant and assess their physical relevance.
It is usually preferred in ML to minimize the number of features and reduce model complexity.
However, this concern does not apply to tree-based algorithms like GBM used in this study because such algorithms only consider features during fitting, which increase the model accuracy.
Consequently, features contributing not or little to the model's accuracy are not selected (for splitting) while the trees are grown \citep{chen2016}, so their impact on model complexity is low \citep{spiliotis2022}.

The features in Tab.~\ref{tab:ra_variables} without ERA5 variable names are so-called engineered features, meaning that they are derived from one or more ERA5 variables.
In the following, we detail the variable selection and the feature engineering.

\paragraph{Shear-related features}
Table~\ref{tab:ra_variables} lists multiple wind-related features aimed at capturing wind shear, i.e., the vertical change of the wind.
Like traditional $C_n^2$ parameterizations, such as W71, we select the friction velocity $u_*$ as the first shear-related feature.
We also include the horizontal wind speed obtained from the zonal and meridional wind fields of ERA5 -- $U_z$ and $V_z$ -- as
\begin{equation}
  M_z = \sqrt{U_z^2 + V_z^2}
  \label{eq:era5_fe_M}
\end{equation}
at $z=10$\,m and $z=100$\,m above ground.
Assuming a power-law wind profile of the form ${M(z) = M_{\text{ref}}\, {(z/z_\text{ref})}^\alpha}$ between the 10\,m and the 100\,m level, we utilize the exponent of the power law $\alpha$ as an additional shear-related feature: 
\begin{equation}
  \alpha(z_1, z_2) = \frac{\log M_{z_2} - \log M_{z_1}}{\log z_2 - \log z_1}.
  \label{eq:era5_fe_alpha}
\end{equation}
The directional shear, i.e., wind turning with height, is included through the absolute angular difference between 10\,m and 100\,m wind direction defined as $\beta = \lvert X_{10} - X_{100} \rvert$ where the wind direction $X_z$ is given as 
\begin{equation}
  X_z = \text{arctan2}\left(-U_z, -V_z\right).
  \label{eq:era5_fe_X}
\end{equation}
The wind direction also serves as a proxy for upstream effects, or fetch, that might influence the observed turbulence.
For example, atmospheric turbulence measured at a station close to the coast can be very different if the wind blows from land to sea or sea to land.
The periodicity of $X_z$ is accounted for by including the sines and cosines of $X_z$ as fetch features instead of $X_z$ directly.

\paragraph{Buyoancy and stability}
The prime candidate to reflect the influence of buoyancy on $C_n^2$ is the sensible heat flux $Q_H$.
Sensible heat flux is also featured in W71 because it captures buoyancy and static atmospheric stability (in ERA5's convention, $Q_H > 0$ during stable/nighttime conditions and $Q_H < 0$ during unstable/daytime conditions). 
Additionally, static stability can be estimated from temperature gradients, so we utilize the absolute values of 2\,m-temperature, skin temperature, and soil temperature, as well as the differences between them as complementary features.

\paragraph{Surface Energy Budget}
The strength of buoyancy depends on the radiation that reaches the surface, so we also include ERA5 radiation fluxes and cloud cover as features to complement the buoyancy information from $Q_H$.
The simplified surface energy budget describes how radiation forces the surface fluxes.
If the small soil heat flux into the ground is neglected, the steady-state energy balance at the surface is \citep{stull1988}
\begin{equation}
  Q_L + Q_H \approx R_{S} + R_{L},
\end{equation}
where $Q_L$ is the latent heat flux due to evaporation and $R_S$ and $R_L$ are the net shortwave and longwave surface radiation fluxes.
The net fluxes $R_\square = R_{\square\downarrow} - R_{\square\uparrow}$ are the difference between incoming/downwelling (index $\downarrow$) and the reflected/upwelling radiation (index $\uparrow$) of shortwave ($\square=S$) and longwave ($\square=L$) radiation.
ERA5 contains the net and downward radiation fluxes as variables. 
Further, it splits the shortwave downwelling radiation $R_{S\downarrow}$ into a direct component $R_{S\downarrow,\text{dir}}$ and a diffuse component $R_{S\downarrow,\text{diff}}$ resulting from scattering by clouds.
Since clouds affect how much shortwave radiation reaches the surface and how much longwave radiation from the surface is reflected, we include ERA5's low cloud cover ($z < \text{ca.\,}2$\,km) and total cloud cover into the dataset.
Close to water bodies, $Q_L$ is expected to be higher due to stronger evaporation leading to lower $Q_H$, and, thus, lower $C_n^2$.
Therefore, $Q_L$ and other moisture-related ERA5 variables are listed as moisture-related features in Tab.~\ref{tab:ra_variables}.

\paragraph{Auxiliary features}
Finally, we include several auxiliary features more loosely related to wind shear and buoyancy, such as boundary layer dissipation rate (BLD), convective available potential energy (CAPE), or the aforementioned gravity dissipation rate (GWD).
Also, certain daily and seasonal patterns exist in meteorology, which we aim to capture through synthetic time-dependent features.
Based on the timestamp of the data point, we compute the sines and cosines (for periodicity) of the normalized hour of the day ($\text{hr}'=2 \pi\, \text{hr}/24$\,h), the normalized day of the year ($\text{day}'=2\pi\, \text{day}/365$\,d), and the normalized month of the year ($\text{month}'=2\pi\,\text{month}/12$\,mo).

\section{Results: feature importance}\label{sec:res_fi}
\begin{figure*}[t]
  \centering
  \includegraphics[width=\textwidth]{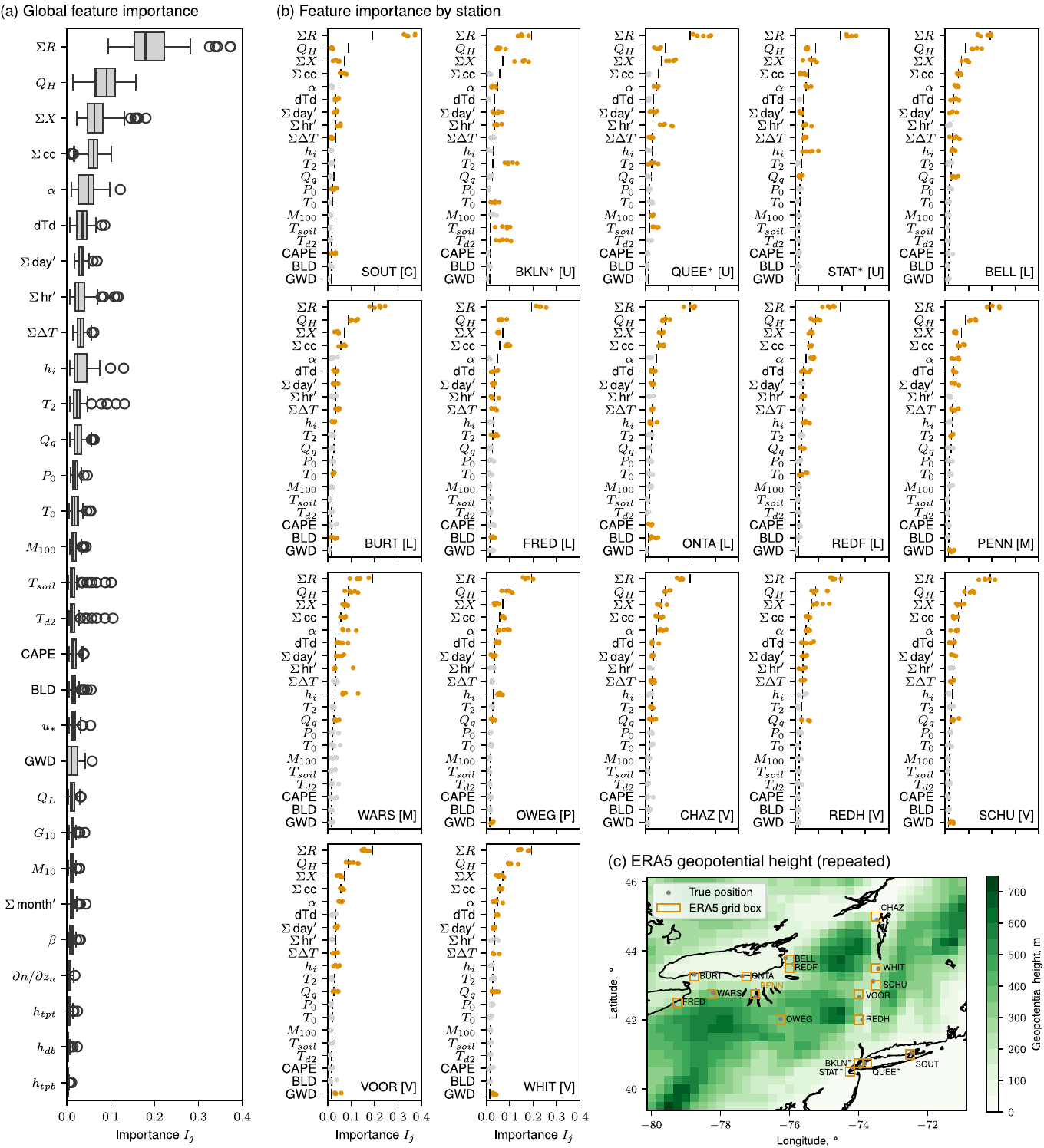}
  \caption{%
    SHAP value-based feature importance (FI) of all OTCliM models (a) aggregated and (b) per training site.
    In (b), urban stations are marked with (*), 
    the top 10 features of each station are highlighted in orange, 
    and the global FI averages are indicated as black dashes.
    Panel (c) repeats the geopotential height from Fig.~\ref{fig:era5_domain} to aid geographical interpretation of the results.
}%
  \label{fig:fi}
\end{figure*}

The physics governing optical turbulence are the same everywhere -- in urban or rural environments or for in-land stations or coastal sites.
The modulating processes are always buoyancy and wind shear.
However, the ERA5 features that best reflect and predict these processes locally can differ between sites.
To assess and quantify potential differences, we present the SHAP-based feature importance values for all trained models in Fig.~\ref{fig:fi}.
To make the FI analysis less verbose, we make use of the linearity of the SHAP values (cf.\ Eq.~\ref{eq:shap_values}) and group the SHAP values of features related to similar physical processes as presented in Tab.~\ref{tab:ra_variables}.
These groups are prefixed with $\Sigma$.

Panel (a) depicts the importance distributions of all features for all models.
This global view reveals that buoyancy-related features such as radiation $\Sigma R$, sensible heat flux $Q_H$, and cloud cover $\Sigma\,\text{cc}$ are key variables for $C_n^2$ prediction.
Also, the shear exponent $\alpha$ is picked up, indicating that the OTCliM models identify wind shear as a modulating factor of turbulence.
The dependency on wind direction $\Sigma X$ suggests that upstream effects influence $C_n^2$ prediction at many sites,
while atmospheric stability is likely captured next to $Q_H$ by the temperature differences $\Sigma \Delta T$ and the boundary layer height $h_i$.
All of the above aligns with our physical knowledge of atmospheric turbulence, indicating that the models picked up physically meaningful relations between ERA5-generated variables and $C_n^2$.
However, the boxplots also show outliers for several features, indicating that not all models agree with the global view.

The FI values are displayed per station in Fig.~\ref{fig:fi}(b), allowing us to assess the outliers in more detail.
The ten features with the highest average FI for each location are highlighted in orange.
Since different stations exhibit different top ten features, 20 features are shown in total, but their markers are colored in light grey if the features are not in the top ten of the site in question.
For reference, the black lines for each feature represent the global average FI values according to panel (a).
Comparing all per-station plots on a high level reveals two key points.
First, although the FI values associated with the five models trained for each site exhibit some scatter, the overall importance of features at one site is consistent between models.
Therefore, we conclude that the models captured features representative of that site throughout the training years.
Secondly, the color coding highlights that different features are important for different sites, suggesting that the processes dominating turbulence vary between locations.

The three urban stations, BKLN, QUEE, and STAT, and the coastal SOUT site deviate most obviously from the mean FI distribution.
In particular, BKLN and QUEE show an above-average dependency of estimated $C_n^2$ on wind direction.
Both stations are located on rooftops and have some tall buildings in their neighborhood \citep{brotzge2020}.
Such inhomogeneous urban conditions are known to influence local measurements \citep{wmo8iii}, so it seems reasonable that turbulence strength observed at BKLN and QUEE depends more strongly on wind direction compared to other more rural locations.
All four stations (urban + SOUT) also show below-average dependency on $Q_H$, which we view as a feature capturing the diurnal cycle.
Instead, the models seem to have picked $\Sigma\, \text{hr}'$ or $h_i$ as predictors for this information.
This behavior differs from all other models, where $Q_H$ is typically the second most important input, in line with more traditional physics-based $C_n^2$ models, such as W71.
This shift, however, need not be viewed as unfavorable but as a demonstration of the power of ML-based modeling to shift from traditional to ``unconventional'' features if complex flow conditions require it.
The downside is that we expect such models to perform more poorly when applied to other sites where the more traditional features are relevant.
Especially, BKLN is expected to generalize poorly because it also depends strongly on the temperature features $T_2$, $T_{d2}$, and $T_{soil}$, deemed irrelevant by almost all other models. 

Less drastic but distinct differences are also visible between the non-urban models.
Wind shear $\alpha$, for example, has lower than average importance assigned for models trained on lake shores (FRED, BURT, ONTA).
The models of another set of stations (WHIT, VOOR, SCHU, PENN, OWEG) picked up the gravity-wave dissipation rate (GWD).
These sites are located in valleys or mountainous areas where gravity waves could modulate near-surface turbulence \citep{lilly1972,palmer1986}.
Still, the dependency is small, and stations CHAZ and REDH, located at the end of valleys but still surrounded by mountains, do not depend on GWD.
A more detailed study of the sites' climatologies would be needed, which is beyond the scope of this work.

Overall, the FI values of most models represent the known physical dependencies of atmospheric turbulence. 
That supports our confidence that our OTCliM approach is well suited for MCP\@.
The ``unconventional'' features picked up by some models are viewed as an advantage of ML-based methods to arrive at accurate predictions even in complex environments.
However, we assume that geographical generalization will be more difficult for such models, which is addressed later in this paper.

\section{Results: OTCliM performance}
After establishing that all OTCliM models picked up physically meaningful dependencies, we turn our analysis to quantifying their performance.
The foundation for this analysis are 85 models trained individually for the 17 NYSM stations and the five training years.
The temporal extrapolation capabilities of each model are quantified in \S\ref{sec:res_mcp} to assess the suitability of OTCliM for MCP.
To put the MCP scores achieved by our approach into perspective, we compare them to the two baseline models, W71 and the in-situ-based GBM models.
Since we have a network of stations available, we also assess the geographical generalizability of the OTCliM models by applying models trained on one site to all other sites.
The results of this cross-site evaluation study are discussed in \S\ref{sec:res_cs}.

\subsection{Temporal extrapolation}\label{sec:res_mcp}
\begin{figure*}[t]
  \centering
  \begin{subfigure}[t]{.495\textwidth}
    \caption{Correlation coefficient $r$ (higher is better)}
    \includegraphics[width=\textwidth]{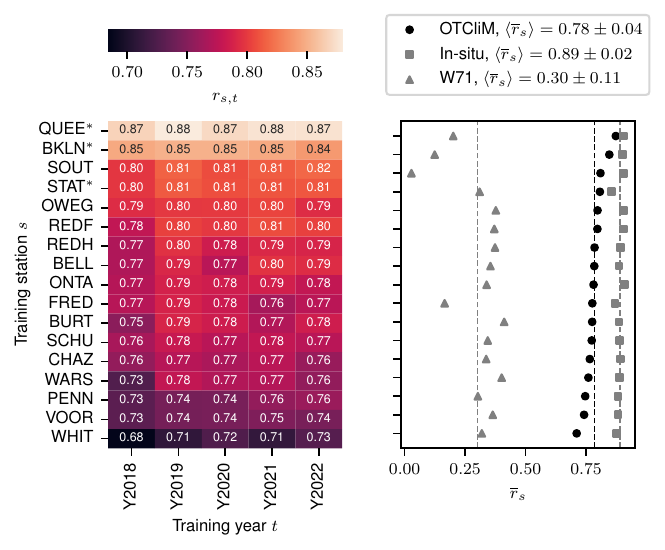}
  \end{subfigure}
  \hfill
  \begin{subfigure}[t]{.495\textwidth}
    \caption{Root-mean-squared error $\epsilon$ (lower is better)}
    \includegraphics[width=\textwidth]{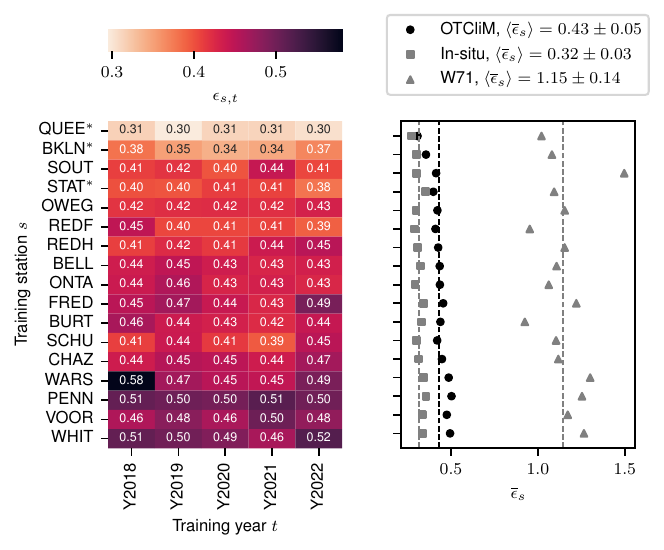}
  \end{subfigure}

  \begin{subfigure}[t]{.495\textwidth}
    \caption{SOUT ($\overline{r}_s = 0.81$, $\overline{\epsilon}_s=0.41$) example predictions}
    \includegraphics[width=\textwidth]{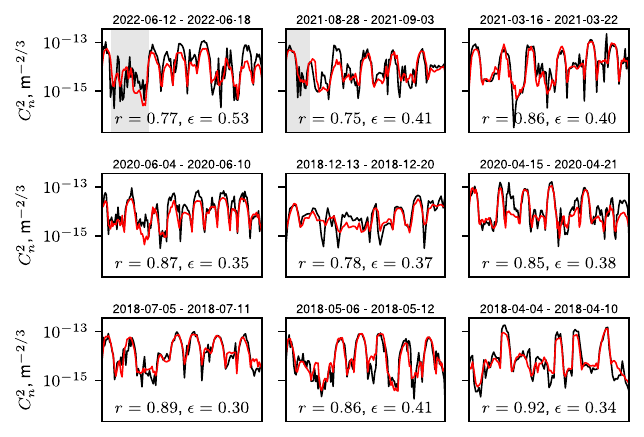}
  \end{subfigure}
  \hfill
  \begin{subfigure}[t]{.495\textwidth}
    \caption{WHIT ($\overline{r}_s = 0.71$, $\overline{\epsilon}_s=0.49$) example predictions}
    \includegraphics[width=\textwidth]{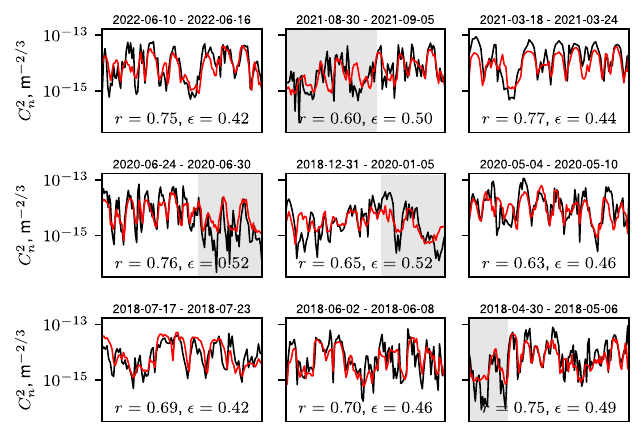}
  \end{subfigure}

  \caption{%
    Performance of OTCliM models compared to the baseline models. 
    The heatmaps show the performance of individual OTCliM models trained with one year of $C_n^2$ observations from site $s$ when evaluated on the four hold-out years of the same site.
    The scatter plots compare the site-averaged performance against the baseline estimates. %
    Seven-day batches of $C_n^2$ are randomly drawn to compare observations (black) against predictions (red) for a high-accuracy (c) and a lower-accuracy OTCliM model (d), with complex turbulence conditions shaded in grey.
  }%
  \label{fig:res_mcp}
\end{figure*}

The temporal extrapolation performance of each model $\hat{f}_{s,t}$ with respect to the correlation coefficient $r$ and the scaled RMSE $\epsilon$ is presented in panels (a) and (b) of Fig.~\ref{fig:res_mcp}, respectively.
The heatmaps for both metrics show the MCP scores $\square_{s,t}$ with their site-averaged values $\overline{\square}_s$ being compared to the baseline models in the accompanying scatter plots.
The heatmap reveals similar patterns for both metrics: 
the score variance across training years (rows) is low for almost all stations, 
while the site-specific performance (columns) varies notably.
In other words, regardless of the year of $C_n^2$ observations used for training, the MCP performance stays consistent, whereas $C_n^2$ at some locations (e.g., QUEE, BKLN, SOUT) is easier to predict from ERA5 than at others (e.g., VOOR, WHIT).
It is unexpected but impressive that the urban models marked with (*) are the highest performing ones due to the typically complex urban climatology, which makes traditional modeling difficult \citep{rotach2005}.
Panels (c) and (d) show a few randomly drawn batches of observed $C_n^2$ with their corresponding predictions.
These curves highlight that the overall performance at SOUT (c) is better than that of WHIT (d) for two reasons.
First, the observed $C_n^2$ (black) at WHIT exhibits more high-frequency oscillations than SOUT, which are missing in the predictions (red).
Second, complex short-duration events seem to occur more frequently for WHIT, which are only partially captured.
Both effects indicated that WHIT is subject to more complex flows than SOUT and that OTCliM misses these details.
We attribute this to missing details in ERA5, as will be discussed later.
For now, we conclude that the models for both locations capture the general trends of $C_n^2$ well and that the observed intra-site performance variation is primarily due to smoothed predictions.
For practical applications such as obtaining a $C_n^2$ climatology, the large-scale behavior of $C_n^2$ is the most relevant, so the smoothing is of limited concern.

The insight that the training year selection has a low influence on the model performance has high practical relevance.
It suggests that models can be trained on archive data and do not necessarily require recent observations if and only if the site's climatology remains comparable between the training period and the envisioned period of model application.
Drastic changes in the station's surroundings, e.g., through construction projects or climate change, would likely break this assumption and, thus, the model's applicability.

The variability across models trained for different stations is likely due to ERA5 not fully capturing all local effects modulating the turbulence.
For example, some landscape features, such as small lakes around PENN and WARS, are smaller than the ERA5 resolution, so they are missed.
The relevance of these local features is highlighted when the OTCliM performance (black circles) is compared to that of the in-situ baseline models in the $\overline{r}_s$ and $\overline{\epsilon}_s$ scatter plots.
While the in-situ models exhibit a small performance scatter around their mean accuracy (dashed lines), a generally widening performance gap is visible, which sets off small (QUEE) and grows toward the lower end of the performance ranking.
Remember that only the input data differs between OTCliM and in-situ (ERA5 vs.\ in-situ observations); the GBM training framework remains the same.
Consequently, the growing gap between the model variants can be attributed to ERA5 holding less predictive power for some locations than others.
Expectedly, the models with lower accuracy (e.g., WHIT, VOOR, PENN) are located close to lakes or in mountainous areas, which are usually subject to complex flows that ERA5 might not well capture.
Surprisingly, the ERA5-based urban models perform almost on par with the in-situ-based models.
This result is impressive, given that urban flows are complex and likely not well represented in ERA5.
Nevertheless, as discussed earlier, GBM seemingly circumvented this issue by shifting from traditional to unconventional features in urban cases.
These features still seem to hold enough predictive power for the high performance observed here.

A final comparison between OTCliM and the traditional W71 parameterization in the scatter plots of panels (a) and (b) shows that OTCliM clearly outperforms W71.
That is because the underlying similarity function (cf.~Eq.~\ref{eq:w71_sim_func}) is empirically determined based on the flow over a flat, unobstructed plain \citep{wyngaard1971} and does not adapt to local topography or climatology.
This limitation becomes especially evident for the urban sites QUEE and BKLN where $\overline{r}_s$ is much below the W71 average (dashed line).
Our OTCliM models, on the other hand, perform well at these locations, highlighting the advantage of accounting for local climatology in complex atmospheric conditions.
In summary, the presented scores demonstrate the capability of our OTCliM approach to accurately extrapolate a 1-year $C_n^2$ time series to multiple years with ERA5 input for a diverse set of locations.

\subsection{Cross-site evaluation}\label{sec:res_cs}
\begin{figure*}[t]
  \centering
  \begin{subfigure}[t]{\textwidth}
    \caption{Relative and absolute correlation coefficient $r$ (higher is better)}
    \centering
    \includegraphics[width=\textwidth]{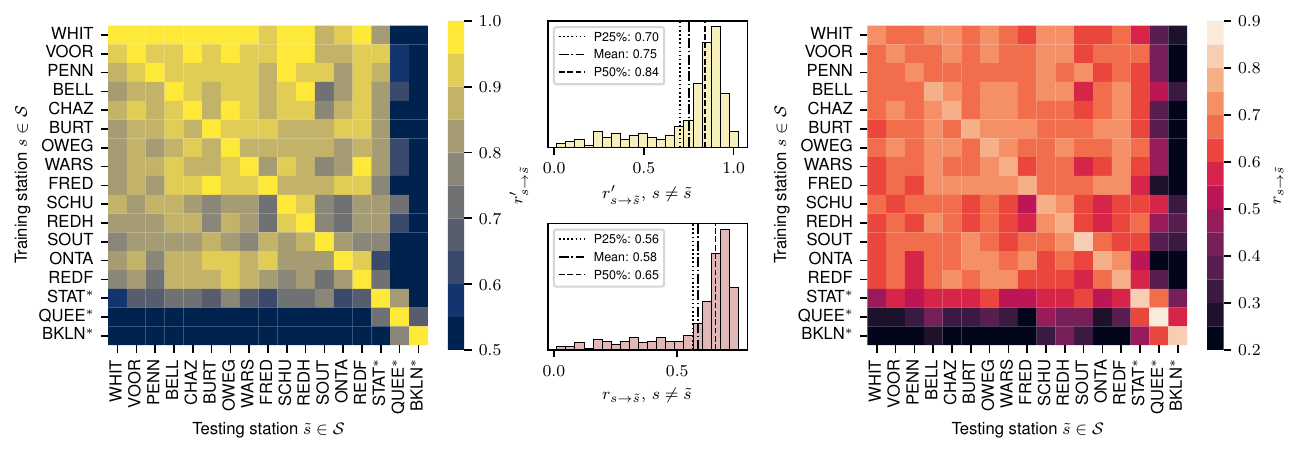}
  \end{subfigure}

  \begin{subfigure}[t]{\textwidth}
    \caption{Relative and absolute root-mean-squared error $\epsilon$ (lower is better)}
    \centering
    \includegraphics[width=\textwidth]{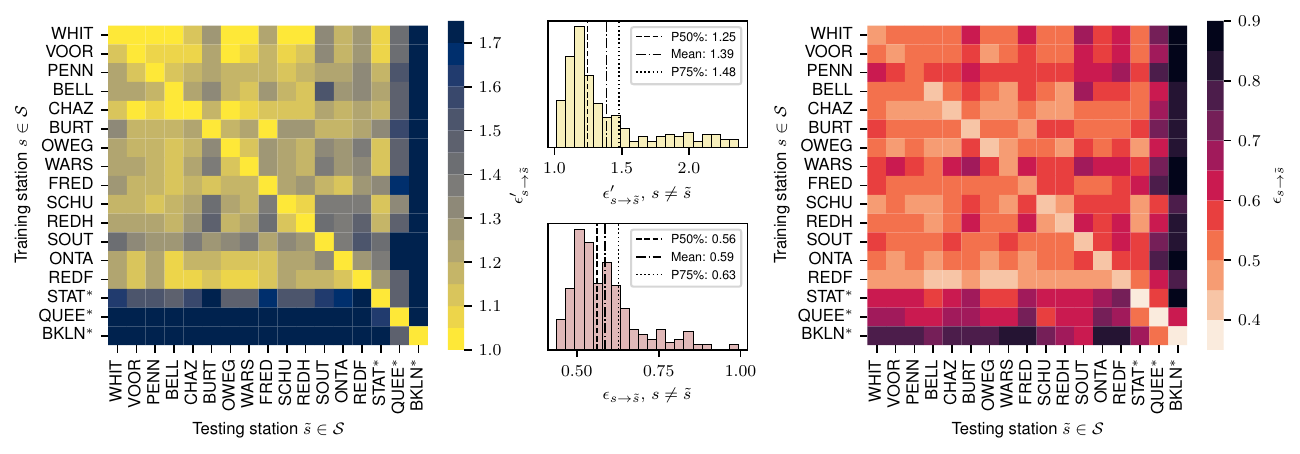}
  \end{subfigure}
  
  \caption{
    Cross-site evaluation performance of OTCliM models. 
    The rows of the heatmaps present the performance of the models trained on site $s$ when evaluated on all other sites $\tilde{s}\in\mathcal{S}$. 
    The left heatmaps show the performance degradation compared to the MCP case, 
    i.e, the relative scores $\square_{s\rightarrow\tilde{s}}'=\square_{s\rightarrow\tilde{s}} / \square_{s\rightarrow{s}}$ with $\square$ representing $r$ or $\epsilon$,
    and the right ones show the absolute scores for reference.
    The histograms depict the distributions of both heatmaps. 
  }%
  \label{fig:cs_full}
\end{figure*}

Next, we assess the geographical generalization capabilities of the OTCliM models by assessing their performance when being evaluated across different sites.
Each model $\hat{f}_{s,t}$ trained on one site $s$ is asked for $C_n^2$ predictions based on five years of ERA5 input data of all other sites ($\tilde{s} \in \mathcal{S} \setminus s$).
The resulting cross-site (c/s) correlation coefficient and RMSE scores are presented in Fig.~\ref{fig:cs_full}.
Each cell $(s, \tilde{s})$ of the heatmaps corresponds to the average score achieved by the five OTCliM models trained on site $s$ (rows) and evaluated on site $\tilde{s}$ (columns).
The heatmaps on the left present the scores normalized by the MCP scores, reflecting the performance degradation of a model when applied to other sites compared to performance on the hold-out years of the original training site. 
The heatmaps on the right show the absolute scores, and the histograms in the middle display the distribution of both relative and absolute scores with diagonals excluded, i.e., without the MCP performance.
The rows and columns of all heatmaps are ordered by mean generalization performance quantified as ${\langle r'_{s\rightarrow\tilde{s}}\rangle}_{\tilde{s}\,\in\,\mathcal{S}\setminus s}$.

Overall, the geographical generalization works very well for the non-urban sites. 
For half of all cases, correlation and RMSE performance degrades by not more than 16\% 
for $r$ and 25\% for $\epsilon$ compared to the original MCP scores.
Considering 75\% of the cases, corresponding to the c/s scores of almost all non-urban sites, the absolute performance falls as low as $r=0.56$ and $\epsilon=0.63$, 
which is still significantly better than the W71 baseline scores of $\langle \overline{r}_s \rangle=0.3$ and $\langle \overline{\epsilon}_s \rangle = 1.15$ (cf.\ Fig~\ref{fig:res_mcp}).
In other words, any non-urban OTCliM model applied to any other non-urban site yields significantly better performance than the traditional approach.

As expected from the FI analysis, the urban models do not perform well in non-urban locations. 
Both relative and absolute c/s scores in Fig.~\ref{fig:cs_full} are low for STAT, QUEE, and BKLN, resulting in long tails in the c/s histograms.
However, the urban models perform better at other urban sites, as indicated by the small square of higher performance in the bottom right corner of the c/s matrices.
It seems that the performance of urban generalization could be linked to the degree of urbanization.
BKLN is the most urban site, QUEE has more vegetation than BKLN, and STAT has more vegetation than QUEE, which is reflected in the progression of c/s performance:
BKLN generalizes most poorly and depends on the most uncommon features, QUEE performs slightly better, and STAT performs even better with FI values similar to the average case.
Nevertheless, this observation might be unique to these New York City stations, which are all relatively close to each other and experience similar mesoscale conditions.

\section{Conclusion}
This study presents OTCliM, a gradient boosting-based measure-correlate-predict approach to obtain climatologies of optical turbulence strength from one year of $C_n^2$ observations and multi-year ERA5 reference data.
A feature importance (FI) analysis based on SHAP values revealed that most models, especially the non-urban ones, learn similar dependencies.
The dominating features reflect the known and expected physical dependencies, suggesting that the OTCliM models captured physically meaningful relations between ERA5 and $C_n^2$.
The FI results also suggest features such as boundary layer dissipation rate or gravity wave dissipation rate as $C_n^2$ dependencies, which are not considered in traditional models.
This result demonstrates how machine learning models could hint at new insights into turbulent processes.
In contrast to most stations, the three urban locations differed significantly from the average FI distribution.
We assume that such models learned very different input-output relations, hindering them from generalizing well to other sites.
In a temporal extrapolation study, we demonstrate that the OTCliM models accurately predict four held-out years of $C_n^2$ from ERA5 inputs across 17 diverse sites in New York State.
The models' performances stay consistent regardless of the year of data selected for training, suggesting archive data can also be used for training.
The trained models are also found suitable for stations located in complex environments such as valleys or cities, which are often difficult to model traditionally.
Minor accuracy variations between models from different sites are visible but can be attributed to local details missing in the coarse ERA5 input data.
For some regions in the world, regional reanalysis datasets are available at higher resolution (e.g., Europe \citep{cerra} or Australia \citep{su2019}) or with more assimilated data (e.g., North America \citep{mesinger2006}), which might help to reduce this problem.
Already with ERA5 input, OTCliM models significantly outperform traditional analytical $C_n^2$ models and achieve scores close to those of GBM models trained on in-situ inputs. 
A geographical generalization study also showed that applying non-urban models to other non-urban sites yields high scores in many cases, indicating that these models generalize well.
The performance only degrades significantly when the urban models are evaluated at non-urban sites.
While these models performed very well in the temporal extrapolation test case, the low scores for cross-site application indicate that the urban models learned very site-specific relations.

The key conclusion for practice from our work is that one year of $C_n^2$ observations is sufficient to obtain reliable site-specific $C_n^2$ climatologies using OTCliM.
Measuring $C_n^2$ at sites of interest for only one year instead of multiple years would free up instruments sooner.
As a result, site survey costs could be reduced, or more locations could be surveyed for the same costs.
OTCliM's high performance in urban environments is highly relevant for FSOC, where terminals can be expected to be located in both rural and urban locations.
The good geographical generalization results of the non-urban models indicate that future OTCliM models trained on a few stations can potentially be used to estimate the spatial distribution of $C_n^2$ over larger areas.
Such $C_n^2$ maps could provide a first indication of locations suitable for FSOC or astronomy with respect to optical turbulence.
However, the region-specific features identified in the FI analysis for lake or mountain environments also suggest that the models must account for local effects.
Developing such multi-region models requires more work before large-scale $C_n^2$ estimates can be considered reliable.
As a first step, users can perform an FI analysis on their trained models and use the results as guidance for expected geographical generalization.
For example, if the models pick up unconventional features, they will likely not generalize to other sites, but if features are more traditional, they might.

Finally, we believe that OTCliM can be highly relevant also for $C_n^2$ forecasts. 
The rapidly advancing ML weather prediction (MLWP) models, such as Graphcast \citep{lam2023} or AIFS \citep{lang2024}, are typically trained on ERA5 and, therefore, produce ERA5-like forecasts.
Currently, not all variables required for OTCliM are available from MLWP, but in principle, OTCliM can be used to translate the MLWP forecasts to $C_n^2$ forecasts.
Similarly, one could train OTCliM on historical data from the Global Forecasting System (GFS) \citep{ncep2015}, a traditional numerical weather prediction system, and convert the GFS forecasts to $C_n^2$ forecasts.

Besides all these advantages, the critical assumption of OTCliM is that the observed year of $C_n^2$ represents the temporal extrapolation range.
If the surroundings change drastically through, e.g., construction, this assumption breaks down, and OTCliM predictions can no longer be considered valid.
Also, our proposed approach currently only predicts near-surface $C_n^2$, but for astronomy and FSOC, higher-level $C_n^2$ or full profiles are also relevant. 
In a previous study, we presented an ML-based framework that can combine $C_n^2$ from multiple levels into one physics-inspired model \citep{pierzyna2023a} for multi-level $C_n^2$ predictions.
How this approach can be integrated into OTCliM remains open for future work.
In summary, our presented OTCliM approach is shown to be highly accurate in diverse meteorological conditions with the potential for geographical generalization.
We believe that OTCliM is a relevant tool for the optical turbulence community for future site surveys and related studies.

\begin{acknowledgments}
This publication is part of the project FREE -- Optical Wireless Superhighways: Free photons (at home and in space) (with project number P19-13) of the research programme TTW-Perspectief, which is (partly) financed by the Dutch Research Council (NWO).
This research is made possible by the New York State (NYS) Mesonet. Original funding for the NYS Mesonet (NYSM) buildup was provided by Federal Emergency Management Agency grant FEMA-4085-DR-NY. The continued operation and maintenance of the NYSM is supported by National Mesonet Program, University at Albany, Federal and private grants, and others.
Sukanta Basu is grateful for financial support from the State University of New York's Empire Innovation Program.
\end{acknowledgments}

\appendix
\section{Ratio of stable to unstable conditions in dataset before and after QA/QC}\label{app:qaqc_stab}

\begin{figure*}[t]
    \centering
    \includegraphics[width=\textwidth]{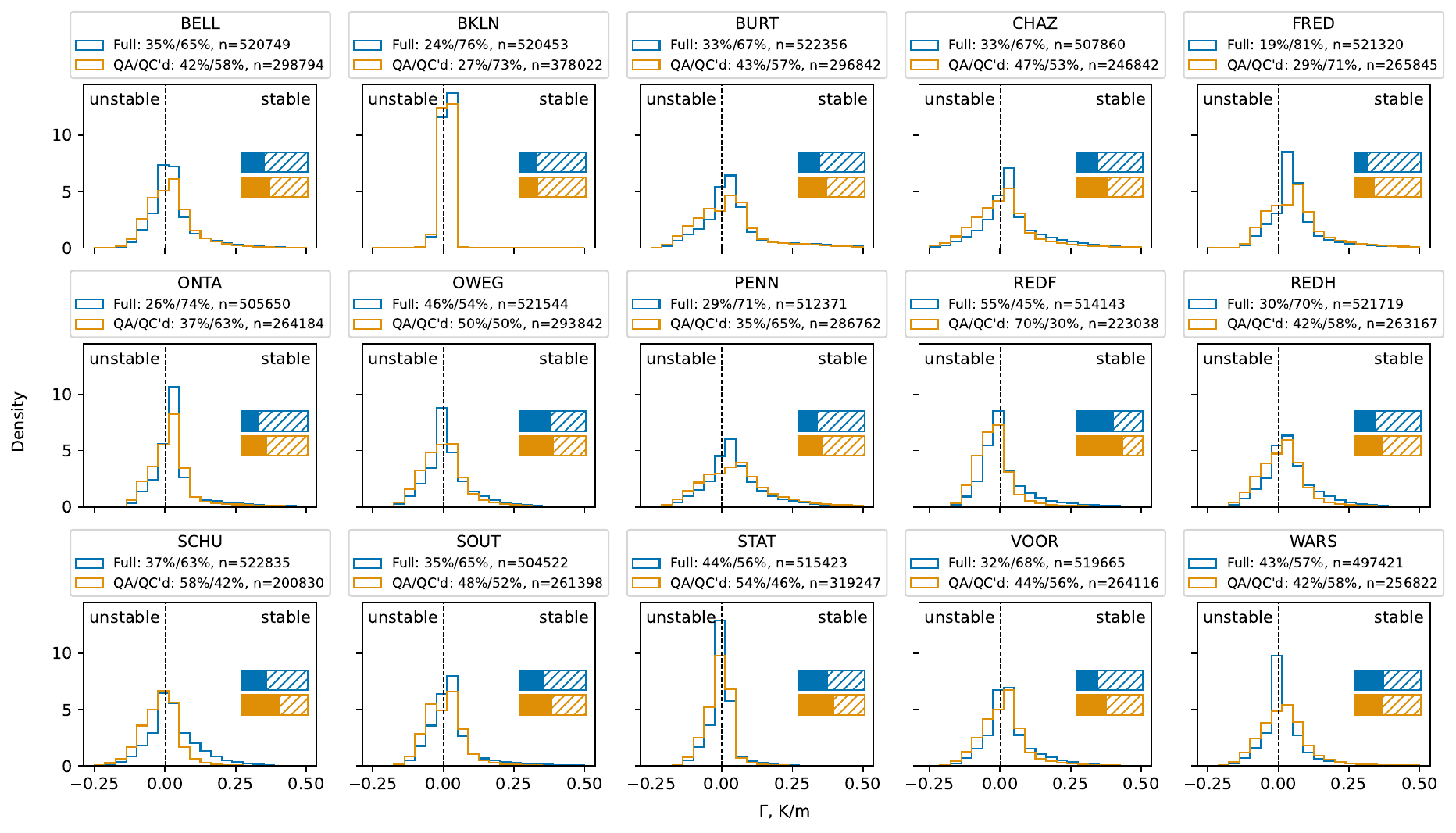}
    \caption{Distribution of bulk potential temperature gradient $\Gamma = (\theta_9 - \theta_2) /7\,\text{m}$ before (blue) and after (orange) applying the quality assurance and control (QA/QC) steps presented in \S\ref{sec:data_nysm_qaqc}. 
    The split between unstable ($\Gamma < 0$, solid bar) and stable ($\Gamma > 0$, hatched bar) atmospheric conditions is visualized by the bar charts in each panel or quantified in the respective legend. The QUEE site is omitted fully because of a malfunctioning instrument.}%
    \label{fig:qa_qc_stability_distr}
\end{figure*}

A three-step quality assurance and control (QA/QC) procedure is described in \S\ref{sec:data_nysm_qaqc}, which aims at filtering unphysical values from our $C_n^2$ dataset to lay the foundation for high-quality OTCliM models.
As discussed in the respective section, the QA/QC procedure discards approximately half of the data points across all stations and all years.
We aim to not significantly change the ratio of unstable to stable atmospheric conditions by QA/QC to expose the OTCliM models to turbulence conditions representative of the respective site.
In Fig.~\ref{fig:qa_qc_stability_distr}, we present the distribution of atmospheric stability before and after QA/QC for each location.
Stability is quantified through the bulk potential temperature gradient $\Gamma = (\theta_9 - \theta_2) /7\,\text{m}$, which is computed from the potential temperature $\theta$ measured by two thermometers at 2\,m and 9\,m above ground \citep{nysmesonet2023}.
These instruments differ from the sonic anemometers used to estimate $C_n^2$. 
They are more reliable, resulting in only 1.5\% of missing values on average due to instrumentation problems compared to 16\% for the sonic anemometers.
The only exception is the QUEE site, which did not have any 9\,m observations during our time of interest, which is why it is omitted in this discussion.
Figure~\ref{fig:qa_qc_stability_distr} shows that the unstable/stable ratio (see bar chart insets) does not change much at most sites due to quality control. 
In general, the number of stable condition samples decreases because stable conditions occur more frequently in winter when instrument malfunctioning due to rain, ice, or snow is more common.
Additionally, the inertial range shrinks with increasing stability and can even disappear \citep{grachev2013}, resulting in more failed fits of the 2/3 slope (QA/QC step one). 
Since the balance between stable and unstable cases remains better than approx.\ 30/70 for all stations, we view our dataset as reasonably well-balanced and adequate for training.

\section{Scaling of $\log_{10} C_n^2$ target data}\label{app:distr_scaling}
\begin{figure*}[t]
  \centering
  \includegraphics[width=.8\textwidth]{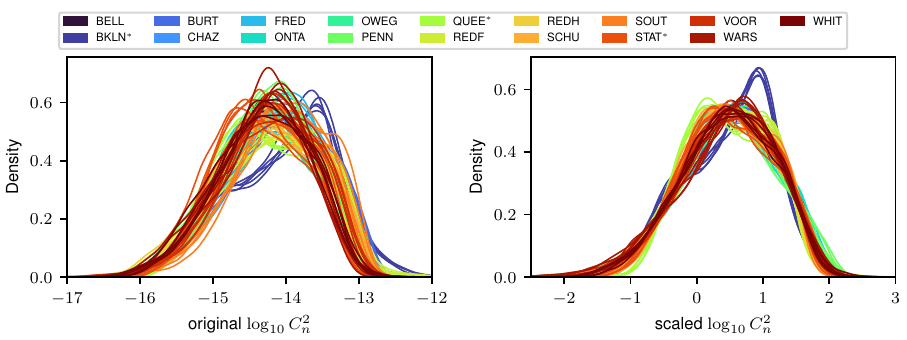}
  \caption{
    Comparison between distributions of unscaled (left) and scaled (right) $\log_{10} C_n^2$.
    Colors indicate the different sites at which $C_n^2$ is observed with urban sites marked by (*).
  }%
  \label{fig:distr_scaling}
\end{figure*}

As described in Sect.~\ref{sec:distr_scaling} of the main text, the $\log_{10} C_n^2$ data are normalized using the 25th and 75th percentiles of the site-specific $\log_{10} C_n^2$ training data (cf.~Eq.~\ref{eq:distr_scaling}).
The effect of this scaling is illustrated in Fig.~\ref{fig:distr_scaling}, which presents the unscaled (left) and scaled (right) data colored by location. 
The plots demonstrate that the scaling makes the range of $\log_{10} C_n^2$ more comparable between sites, which is crucial to obtain comparable performance and FI scores.
That is especially important for the urban BKLN site (dark blue), where the unscaled right tail of the distribution is shifted to the right by half an order of magnitude (higher OT strength) compared to the other sites.
To keep the scaling robust and general (i.e., applicable to non-gaussian distributions), we scale using the inter-quartile range and not the commonly used standard deviation. 

\section{Details about baseline models}\label{app:baseline}
This section details how the traditional W71 $C_n^2$ parameterization and the in-situ-based GBM models are setup and utilized as performance baselines for our proposed OTCliM approach.

\subsection{W71 $C_n^2$ parameterization}
All variables on which the W71 equations (cf.~Eqns.~\ref{eq:w71} -- \ref{eq:w71_sim_func}) depend are available from ERA5 (cf.\ Tab.~\ref{tab:ra_variables}).
Only the dynamical sensible heat flux $Q_H$ from ERA5 needs to be converted to its kinematic form $\overline{w'\theta'}$ as $\overline{w'\theta'} = - Q_H / (\rho\, c_p)$, 
where $\rho$ and $c_p$ are the density and specific heat capacity of air \citep{stull1988}.
Also, the sign of $Q_H$ is flipped because ERA5 treats an upward $Q_H$ as negative, whereas W71 expects the upward $\overline{w'\theta'}$ to be positive.
After this conversion, we utilize the same ERA5 data extracted for the 17 NYSM stations that we use to train the OTCliM models to evaluate the W71 equations. 
The results are 17 five-year $C_n^2$ time series, which can be compared to the observed $C_n^2$ evolution at each station.
The corresponding RMSE and $r$ scores are presented in Fig.~\ref{fig:res_mcp}.

\subsection{In-situ GBM models}
These upper-bound models are similar to the OTCliM models but use different input data.
The OTCliM model of a specific site $s$ is a GBM model (cf.~\ref{sec:meth_gb}) that employs ERA5 input data extracted from the grid box containing site $s$ and $C_n^2$ observed at site $s$.
Instead of ERA5, one can utilize in-situ weather data observed at $s$.
The resulting GBM in-situ models are quite similar, for example, to work presented by \citet{wang2016}, \citet{jellen2021}, or \citet{pierzyna2023a}.
As for selecting ERA5 variables (cf.~\ref{sec:data_era5}), we aim to capture buoyancy and wind shear with multiple features based on the instruments deployed at the NYSM stations.
As buoyancy features, 
we compute the temperature gradient $\Gamma = (T_9 - T_2) / (9\text{\,m} - 2\text{\,m})$ from observed 2\,m and 9\,m temperature, 
and utilize the observed 30\,min sensible heat flux and observed incoming radiation.
Assuming the velocity at the ground to be close to zero, we obtain a crude estimate of the bulk wind shear $S=M_{10} / 10\text{\,m}$ where $M_{10}$ is the horizontal windspeed at 10\,m.
Additionally, observed friction velocity $u_*$, latent heat flux $\overline{w'q'}$, and the dew point spread $\Delta T_d = T_2 - T_{d2}$ are included as features, where $T_{d2}$ is the 2\,m dew point temperature.
To account for atmospheric stability, $\Gamma$ and $S$ are combined into a bulk Richardson number as $\text{Ri}=(g/\overline{T})\, \Gamma / S^2$ with $g=9.81$\,m\,s$^{-2}$.
Upstream effects are captured by adding sine and cosine of the 10\,m wind direction.
All these variables form the in-situ dataset, which is used to train GBM models in the same fashion as OTCliM: 
each in-situ model is trained on one year of observations from site $s$ and evaluated on the hold-out years of the same site.
The resulting scores are presented as upper baseline in Fig.~\ref{fig:res_mcp}.

\bibliography{references}

\end{document}

%% file: nysm_names.tex
\begin{tabular}{llrrl}
\toprule
ID & Nearest City & Latitude, $^{\circ}$\,N & Longitude, $^{\circ}$\,W & Environment \\
\midrule
BELL & Belleville & 43.79 & 76.11 & Lake \\
BKLN$^{*}$ & Brooklyn & 40.63 & 73.95 & Coastal / Urban \\
BURT & Burt & 43.32 & 78.75 & Lake \\
CHAZ & Chazy & 44.90 & 73.46 & Valley \\
FRED & Fredonia & 42.42 & 79.37 & Lake \\
ONTA & Ontario & 43.26 & 77.37 & Lake \\
OWEG & Owego & 42.03 & 76.26 & Plateau \\
PENN & Penn Yan & 42.66 & 76.99 & Mountainous / Lake \\
QUEE$^{*}$ & Queens & 40.73 & 73.82 & Coastal / Urban \\
REDF & Redfield & 43.62 & 75.88 & Lake \\
REDH & Red Hook & 42.00 & 73.88 & Valley \\
SCHU & Schuylerville & 43.12 & 73.58 & Valley \\
SOUT & Southold & 41.04 & 72.47 & Coastal \\
STAT$^{*}$ & Staten Island & 40.60 & 74.15 & Coastal / Urban \\
VOOR & Voorheesville & 42.65 & 73.98 & Valley \\
WARS & Warsaw & 42.78 & 78.21 & Mountainous / Lake \\
WHIT & Whitehall & 43.49 & 73.42 & Valley \\
\bottomrule
\end{tabular}

%% file: era5_vars.tex
\begin{tabular}{llllll}
\toprule
Category & Symbol & Grouping & ERA5 variable name & Description & Unit \\
\midrule
Wind & $M_{100}$ &   & - & 100m horizontal wind speed, see eq.~\ref{eq:era5_fe_M} & m\,s$^{-1}$ \\
Wind & $M_{10}$ &   & - & 10m horizontal wind speed, see eq.~\ref{eq:era5_fe_M} & m\,s$^{-1}$ \\
Wind & $G_{10}$ &   & i10fg & Instantaneous 10m wind gust & m\,s$^{-1}$ \\
Shear & $u_*$ &   & zust & Friction velocity & m\,s$^{-1}$ \\
Shear & $\alpha$ &   & - & Shear coefficient of power-law wind profile, see eq.~\ref{eq:era5_fe_alpha} & - \\
Shear & $\beta$ &   & - & Directional shear, $\lvert X_{10} - X_{100} \rvert$ & $^{\circ}$ \\
Fetch & $\cos\left(X_{100}\right)$ & $\Sigma X$ & - & Cosine of 100m wind direction, see eq.~\ref{eq:era5_fe_X} & - \\
Fetch & $\cos\left(X_{10}\right)$ & $\Sigma X$ & - & Cosine of 10m wind direction, see eq.~\ref{eq:era5_fe_X} & - \\
Fetch & $\sin\left(X_{100}\right)$ & $\Sigma X$ & - & Sine of 100m wind direction, see eq.~\ref{eq:era5_fe_X} & - \\
Fetch & $\sin\left(X_{10}\right)$ & $\Sigma X$ & - & Sine of 10m wind direction, see eq.~\ref{eq:era5_fe_X} & - \\
Buoyancy & $Q_H$ &   & ishf & Instantaneous surface sensible heat flux & W\,m$^{-2}$ \\
Temperature & $T_2$ &   & t2m & 2m temperature & K \\
Temperature & $T_0$ &   & skt & Skin temperature & K \\
Temperature & $T_{\text{soil}}$ &   & stl1 & Soil temperature & K \\
Stability & $\Delta T_0$ & $\Sigma \Delta T$ & - & Temperature difference $T_{\text{soil}} - T_0$ & K \\
Stability & $\Delta T_1$ & $\Sigma \Delta T$ & - & Temperature difference $T_0 - T_2$ & K \\
Radiation & $R_{S\downarrow,\text{diff}}$ & $\Sigma\, R$ & msdwswrf & Mean diffuse downwelling short-wave radiation flux at surface  & W\,m$^{-2}$ \\
Radiation & $R_{S\downarrow,\text{dir}}$ & $\Sigma\, R$ & msdrswrf & Mean direct downwelling short-wave radiation flux at surface & W\,m$^{-2}$ \\
Radiation & $R_S$ & $\Sigma\, R$ & msnswrf & Mean net short-wave radiation flux at surface & W\,m$^{-2}$ \\
Radiation & $R_{L\downarrow}$ & $\Sigma\, R$ & msdwlwrf & Mean downwelling long-wave radiation flux at surface & W\,m$^{-2}$ \\
Radiation & $R_L$ & $\Sigma\, R$ & msnlwrf & Mean net long-wave radiation flux at surface & W\,m$^{-2}$ \\
Cloud & lcc & $\Sigma\, \text{cc}$ & lcc & Low cloud cover fraction (0 - 1) & - \\
Cloud & tcc & $\Sigma\, \text{cc}$ & tcc & Total cloud cover fraction (0 - 1) & - \\
Moisture & $\Delta T_d$ &   & - & Dew point spread, $T_{d2} - T_2$ & K \\
Moisture & $T_{d2}$ &   & d2m & 2m dew point temperature & K \\
Moisture & $Q_q$ &   & ie & Instantaneous moisture flux & kg\,m$^{-2}$\,s$^{-1}$ \\
Moisture & $Q_L$ &   & mslhf & Mean surface latent heat flux & W\,m$^{-2}$ \\
Auxilary & $h_i$ &   & blh & Boundary layer height & m \\
Auxilary & BLD &   & bld & Boundary layer dissipation rate & J\,m$^{-2}$ \\
Auxilary & GWD &   & gwd & Gravity wave dissipation & J\,m$^{-2}$ \\
Auxilary & CAPE &   & cape & Convective available potential energy & J\,kg$^{-1}$ \\
Auxilary & $P_0$ &   & msl & Mean sea level pressure & Pa \\
Auxilary & $\partial n/\partial z_a$ &   & dndza & Mean vertical gradient of refractivity inside trapping layer & m$^{-1}$ \\
Auxilary & $h_{db}$ &   & dctb & Duct base height & m \\
Auxilary & $h_{tpb}$ &   & tplb & Trapping layer base height & m \\
Auxilary & $h_{tpt}$ &   & tplt & Trapping layer top height & m \\
Time & $\cos \text{hr}'$ & $\Sigma\, \text{hr}'$ & - & Cosine of normalized hour of the day & - \\
Time & $\sin \text{hr}'$ & $\Sigma\, \text{hr}'$ & - & Sine of normalized hour of the day & - \\
Time & $\cos \text{day}'$ & $\Sigma\, \text{day}'$ & - & Cos of normalized day of the year & - \\
Time & $\sin \text{day}'$ & $\Sigma\, \text{day}'$ & - & Sin of normalized day of the year & - \\
Time & $\cos \text{month}'$ & $\Sigma\, \text{month}'$ & - & Cos of normalized month of the year & - \\
Time & $\sin \text{month}'$ & $\Sigma\, \text{month}'$ & - & Sin of normalized month of the year & - \\
\bottomrule
\end{tabular}